\documentclass[conference]{IEEEtran}
\IEEEoverridecommandlockouts
\usepackage{cite}
\usepackage{amsmath,amssymb,amsfonts}
\usepackage{algorithmic}
\usepackage{graphicx}
\usepackage{textcomp}
\usepackage{xcolor}
\usepackage{bbm}
\usepackage{mathtools}

\usepackage{makecell}
\usepackage{booktabs}
\usepackage{caption}
\usepackage{subcaption}
\usepackage{listings}
\usepackage[disable]{todonotes}
\usepackage[hyperfootnotes=false, colorlinks=true, linkcolor={blue!50!black}, citecolor={blue!50!black},urlcolor={black}]{hyperref} 
\usepackage{balance}

\usepackage[inline]{enumitem}

\usepackage{authblk}

\newcommand{\ra}[1]{\renewcommand{\arraystretch}{#1}}
\usepackage{arydshln}
\usepackage{float}

\def\BibTeX{{\rm B\kern-.05em{\sc i\kern-.025em b}\kern-.08em
    T\kern-.1667em\lower.7ex\hbox{E}\kern-.125emX}}
\begin{document}

\title{Magpie: Automatically Tuning Static Parameters for Distributed File Systems using Deep Reinforcement Learning}

\makeatletter
\newcommand{\linebreakand}{%
  \end{@IEEEauthorhalign}
  \hfill\mbox{}\par
  \mbox{}\hfill\begin{@IEEEauthorhalign}
}
\makeatother

\author[1]{Houkun Zhu}
\author[1]{Dominik Scheinert}
\author[2]{Lauritz Thamsen}
\author[3]{Kordian Gontarska}
\author[1]{Odej Kao}
\affil[1]{Technische Universit{\"a}t Berlin, Berlin, Germany}
\affil[2]{University of Glasgow, Glasgow, United Kingdom}
\affil[3]{HPI, University of Potsdam, Potsdam, Germany}
\affil[ ]{Emails: \{h.zhu, firstname.lastname\}@tu-berlin.de, lauritz.thamsen@glasgow.ac.uk, kordian.gontarska@hpi.de}

\maketitle

\begin{abstract}
Distributed file systems are widely used nowadays, yet using their default configurations is often not optimal. At the same time, tuning configuration parameters is typically challenging and time-consuming.
It demands expertise and tuning operations can also be expensive. This is especially the case for static parameters, where changes take effect only after a restart of the system or workloads.

We propose a novel approach, Magpie, which utilizes deep reinforcement learning to tune static parameters by strategically exploring and exploiting configuration parameter spaces. To boost the tuning of the static parameters, our method employs both server and client metrics of distributed file systems to understand the relationship between static parameters and performance. Our empirical evaluation results show that Magpie can noticeably improve the performance of the distributed file system Lustre, where our approach on average achieves 91.8\% throughput gains against default configuration after tuning towards single performance indicator optimization, while it reaches 39.7\% more throughput gains against the baseline.
\end{abstract}

\begin{IEEEkeywords}
performance optimization, parameter tuning, reinforcement learning, distributed storage systems, cluster configuration
\end{IEEEkeywords}

\section{Introduction}
Distributed storage systems play a significant role in computer clusters. Especially in the context of big data applications, a large volume of data needs to be stored reliably. To satisfy the storage and analysis demand, distributed file systems (DFSs) like Lustre\cite{braam2019lustre} and Ceph\cite{weil2007ceph} are used. They provide flexibility, high availability, and low cost to users. For example, to improve availability and access speed, users can increase the replication of data in such DFSs. However, as the default settings of such systems are often not optimal and tuning parameters can bring vast performance gains\cite{zadok2015parametric, cao2018towards, cao2020carver}, finding the near-optimal settings is desired.

However, parameter tuning is challenging owing to a large number of parameters, non-linear system behaviors, and dependencies among parameters\cite{lyu2020sapphire,herodotou_survey_2021,li2017capes,zhang2019end}. Many parameter configurations can be explored without restarting DFS workloads or systems. For instance, most DFSs have a parameter to control the maximum concurrent client requests, and changes take effect immediately after changing the parameter value. At the same time, there are also many essential \emph{static parameters}, which also need to be tuned carefully but changing those parameters takes effect only after either a workload or the DFS is restarted. For example, the parameter that controls the number of service threads, which often affects the write and read performance, is only changed effectively after restarting the DFS. Consequently, tuning static parameters is even more challenging due to the restarting cost of the DFS or workload. It is, therefore, essential to tune static parameters with few tuning trials.

As manual tuning is demanding to users, many automatic parameter tuning systems have been proposed~\cite{zadok2015parametric, cao2019practical,scheinert2021enel,cheng2021aioc2,huang2019automatic,young2015optimizing,chen2015machine,lyu2020sapphire, lorena2008evolutionary,yigitbasi2013towards, zhang2021convolutional, berkenkamp2021bayesian, wu2019hyperparameter, dalibard2017boat, alipourfard2017cherrypick, saboori2008autotuning, zhang2019end, li2017capes, jomaa2019hyp, neary2018automatic, van2017automatic, zhu2017bestconfig, zhang2021hbox, li2019qtune}.
Specifically, previous works on DFS parameter tuning focus only on dynamic parameters and cannot be applied to static parameters tuning directly. For example, CAPES~\cite{li2017capes} uses Deep Reinforcement Learning (DRL) to model the behavior of DFS to discover the optimal configurations. As it can only change the value of a single parameter within a fixed step size, the requirement of scarce tuning for optimizing static parameters is not met by CAPES. 
General tuning approaches, on the other hand, typically treat DFSs as black boxes and cannot use server and client load information during tuning. It, therefore, demands more profiling data to understand the behavior of DFS. For instance, BestConfig\cite{zhu2017bestconfig} searches the parameter space iteratively to optimize performance. Its initial sampling of parameter space is essential for its final configurations recommendation. This reliance on extensive profiling or, else, a good initial sampling impedes its applicability for static parameter tuning.

We identify the following challenges in DFSs static parameters tuning:
\begin{enumerate*}[series = tobecont, label=(\arabic*)]

	\item Workloads are often dynamic~\cite{cao2018towards} and thus, the tuning of static parameters also needs to be adaptive to workload changes.
	\item There are regularly multiple objectives that need to be optimized simultaneously. For instance, one may need to improve latency and throughput in parallel.
	\item Tuning static parameters is more expensive and challenging as specific workloads or entire systems need to be restarted to apply changes and thus we cannot explore parameter space thoroughly.
\end{enumerate*}

\begin{figure*}[h]
	\centering
    \includegraphics[width=\linewidth]{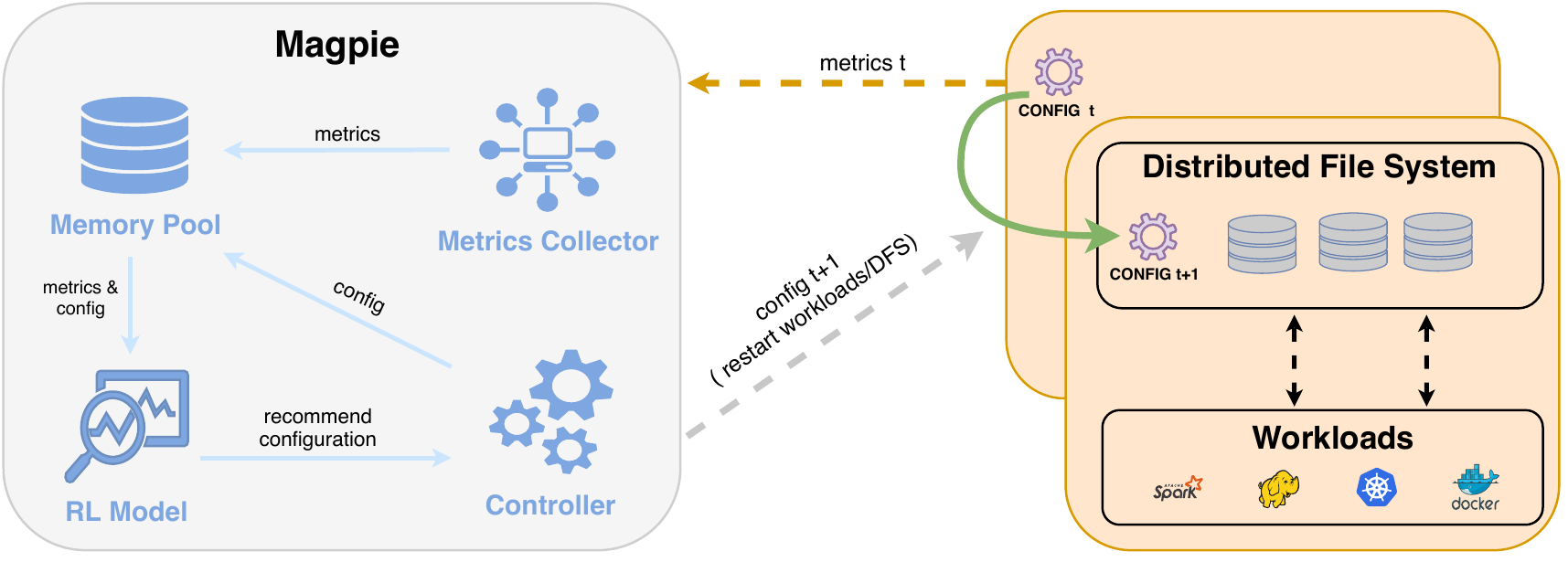}
	\caption{General idea of Magpie. The left side presents the four main components of Magpie and the right side illustrates the distributed file system (DFS) and workloads running on it. 
	Magpie collects metrics for a given configuration $t$ and then analyzes its execution history to recommend a new configuration $t+1$ to the DFS.
	Both the DFS and / or the workloads will then be restarted to apply new configurations.}
	\label{fig:magpie_dfs_interaction}
\end{figure*}
To address those challenges, we propose Magpie, a static parameter tuning system for storage performance optimization using DRL. Our approach employs DRL to explore and exploit parameter space strategically. Specifically, after evaluating the new configurations according to performance indicators we collected from the DFS, it tunes the more promising parameters as indicated by changes in performance metrics. Additionally, our method takes server and client resource usage (e.g., CPU, RAM, cache usage) into account to better understand the parameters' influence on the system. As our model learns the relation between configuration parameters and the system performance, it continuously adapts to workload changes. Furthermore, Magpie adopts scalarization to tune multiple parameters in parallel.

In this paper, we make the following contributions:
\begin{enumerate}
    \item A novel approach to tune the static parameters of DFS using Deep Reinforcement Learning. Both client and server resource usage of DFS are used to boost tuning, while the approach supports tuning towards multiple objectives in parallel.
    \item A prototype of our approach is implemented and shared in a repository\footnote{\url{https://github.com/dos-group/magpie}}, accompanied by the documentation regarding environment setup, system configuration, and experiment execution.
    \item A set of empirical evaluation results that show that our approach can improve DFS performance significantly with a limited number of samples, suitable for tuning static parameters of a state-of-the-art DFS running different workloads.
\end{enumerate}

\section{Approach}

Magpie continuously tunes static parameters of a DFS using deep reinforcement learning. 
By exploring and exploiting the parameter space strategically, the performance optimization of a storage system is possible even in the absence of representative historical data to learn from.
In accordance with standard RL terminology, the DFS and its workloads are the environment of our agent, which is also illustrated in~\autoref{fig:magpie_dfs_interaction}.
The performance indicators of the DFS are then collected by the Metrics Collector and stored in the Memory Pool. 
The RL model analyzes the previous tuning history and the corresponding observed impacts on performance.
Taking the current DFS status into account, it then 
recommends a new configuration, which is applied to the DFS by the Controller and stored in the Memory Pool.
As indicated before, this allows the RL model to later learn and update its policy.

\subsection{Problem Definition}
\label{sec:probelm_definition}

The objective of a performance tuning system $A: \mathbbm{F},\Lambda \rightarrow \lambda$ is to find the best parameter $\lambda$ in the parameter space $\Lambda$ which optimizes the performance $P$ of the DFS for workload F over the set of workloads $\mathbbm{F}$\cite{jomaa2019hyp}.
The m-dimensional parameter space $\Lambda$ is defined as $\Lambda = \lambda_1 \times \lambda_2 \times \lambda_3 \times ... \times \lambda_m$, in which parameters can be continuous or discrete values. Since categorical parameters can be mapped to discrete parameters, this yields $\forall i \in \{1,...,m\}: \lambda_i \in \mathbbm{R}\ \lor\ \lambda_i \in \mathbbm{Z}$, i.e., all parameters are either continuous or discrete. Furthermore, because parameters are often bounded in distributed file systems, we express them by adding a set of constraints $C = \{C_1, C_2, ..., C_n\}$, with each constraint $C_i$ defined as
\begin{align*}
 C_i \coloneqq \lambda_j  \oplus B_i,
\end{align*}
where $ \oplus \in \{<, \leq, \geq, >\}$ is a comparison operator, $j\in \{1,...,m\}$ denotes a bounded parameter, and $B_i \in \mathbbm{R}$ is the corresponding boundary.
 
The performance of a DFS is oftentimes estimated using a set of metrics.
Thus, we define performance as $P = P_1 \times P_2 \times P_3 \times ... \times P_k$ in $k$-dimensional space, i.e., we consider $k$ different metrics.
In general, one may need to optimize many performance metrics at the same time. To address this multi-objective optimization issue, one usually refers to either
\begin{enumerate*}[series = tobecont, label=(\arabic*)]
\item Scalarization, where a multi-objective optimization problem is converted to a single-objective one, or 
\item Multi-Policy Searches, which can provide solutions that are Pareto dominated but not comparable to each other~\cite{van2014multi}.
\end{enumerate*}
We employ the former in our approach, i.e., scalarization, as our goal is to build an end-to-end automatic tuning system, without the need to manually choose from a potential set of Pareto optimal solutions. We use linear-scalarization, i.e., a weighted sum
\begin{equation*}
\sum_{i=1}^{s} w_{i}\times norm(P_{i}),
\end{equation*}

where $norm(\cdot)$ stands for a normalization function, and $w_i$ indicates the weight for parameter i. Normalization is used to standardize the scale for different metrics. For example, one can use min-max normalization. Weights are used to specify the tuning preference. For instance, if one aims to optimize throughput and IOPS with the same preference, they can set the same weight value, e.g., $w_1=w_2=1$.

The aim of a static parameter tuning system is to find the parameter $\lambda^*$ for workload $F$ such that the performance objective function $G$ is maximized under given parameter constraints, which can be represented as:
\begin{align*}
	\lambda^* &= argmax_{\lambda \in \Lambda} G(P) \quad s.t \quad C\\
	 &= argmax_{\lambda \in \Lambda} \sum_{i=1}^{s} w_{i} \times norm(P_{i}) \quad s.t \quad C.
\end{align*}
With this function, we search for the optimal parameter.

\subsection{Problem Representation} %
\label{subsec:problem_representation}
In order to tune static parameters using reinforcement learning, it is required to model the problem as a Markov Decision Process.
\begin{figure}
    \centering
    \includegraphics{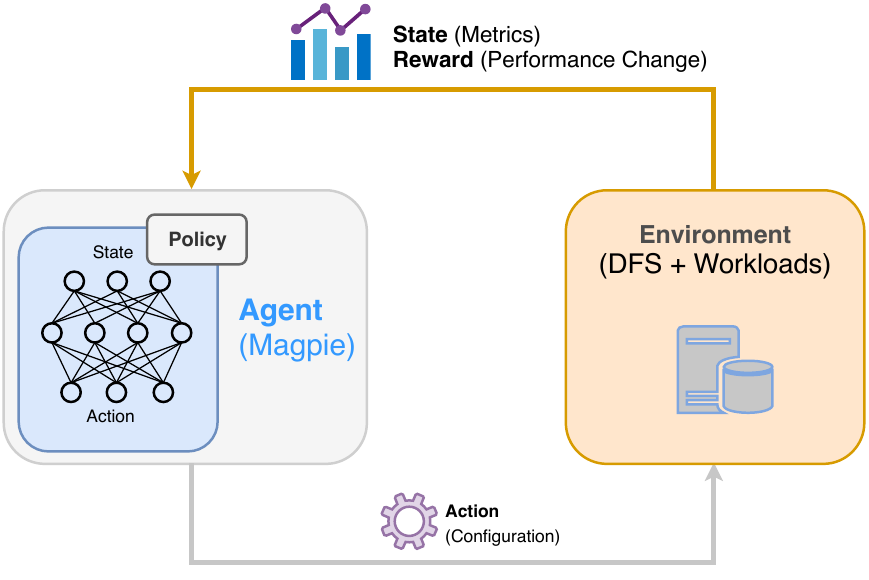}
    \caption{RL problem representation. Each component is illustrated with its role. 
    The agent (Magpie) interacts with the environment (DFS + workloads) by collecting the state (metrics) and reward (performance gains) and taking actions (applying new configurations).}
    \label{fig:problem_representation}
\end{figure}
As shown in~\autoref{fig:problem_representation}, the main components are Magpie (Agent) and a target DFS with its workload(s) (Environment). 
Magpie is interacting with the DFS by tuning different knobs and observing the performance change to understand the relation behind it. 
In the following, the various components are defined in more detail.

\subsubsection{Agent}
Magpie is the agent in this problem. 
It collects metrics of a DFS and then recommends a new configuration to the DFS to improve its performance.

\subsubsection{Environment}
The environment of our problem is a DFS with its workload(s). 
The agent (Magpie) interacts with the environment to discover the optimal configuration for a DFS.

\subsubsection{State}
The state is one of the key elements in reinforcement learning and describes the current condition of the environment. 
We use metrics of the DFS to represent its state, which are in varying units and scales. 
To better interpret different metrics, each metric value is normalized to $[0,1]$ using a normalization function $norm(\cdot)$ that applies min-max normalization. One can also utilize other normalization functions.
The respective boundaries can either be derived using domain knowledge, or inferred from provided data.
Therefore, the state $s$ at step $t$ can be represented as
\begin{align*}
    s_t &= \begin{bmatrix} s_t(1), s_t(2),..., s_t(k)\end{bmatrix}\\
        &=\begin{bmatrix} norm(P_1), norm(P_2),... , norm(P_k) \end{bmatrix}.
\end{align*}

\subsubsection{Action}
A specific new configuration recommended by Magpie is called action. 
For step $t$, an action $a$ is defined as
\begin{align*}
        a_t &= [a_t(1), a_t(2), ..., a_t(m)]\\
        &=[\lambda_1, \lambda_2, ..., \lambda_m].
\end{align*}

In each step, Magpie utilizes its latest policy to perform an action (recommend a new configuration) to the DFS. 
Unlike many other DFS tuning systems, Magpie can tune all parameter values simultaneously at a single step.

\subsubsection{Reward}
In order to inform the agent whether a specific recommended action for a particular state has been useful or shall be avoided in the future, the concept of reward is needed. 
In our context, the reward is the proportional performance change between the current and the previous state.
In Magpie, we collect performance indicators directly from the server-side, to have a consistent and transparent tuning.
Reward at step $t$ can be represented as 
\begin{equation*}
    r_t =  \frac{\sum_{i=1}^{k} w_i s_{t+1}(i) -\sum_{i=1}^{k}w_i s_t(i)}{\sum_{i=1}^{k}w_i s_t(i)}.
\end{equation*}

\subsubsection{Policy}
The policy is realizing the mapping $\mu_t : s_t \rightarrow a_t$ from state to action. 
It can be stochastic or deterministic. 
For Magpie, we use a deterministic policy because of its low sample complexity, so that we can learn a policy with a limited number of tuning steps.
Since in DRL a policy is often implemented by a neural network, accordingly, we would then use the parameter $\theta$ (weights and bias of the neural network) to parametrize the policy function as $\mu_\theta$.

\subsection{DDPG in Magpie} %
\label{subsec:ddpg_in_magpie}
The parameter space for DFS is typically continuous. 
This makes it on the one hand difficult to apply the commonly used Q learning due to the global maximization of Q. On the other hand, discretizing continuous parameters could result in excluding optimal configuration even before optimization.
Hence, we use Deep Deterministic Policy Gradient (DDPG), which can directly and quickly learn promising actions in high-dimensional continuous space.
DDPG is an actor-critic RL algorithm, with both actor and critic being realized as individual neural network models.
As shown in~\autoref{fig:ddpg_network}, the actor recommends a new configuration to a DFS while the critic evaluates the actor's recommendation and gives the actor feedback.
Following this, the procedure of applying DDPG in Magpie can be separated into the two parts, learning and acting.

\begin{figure}[h]
	\includegraphics{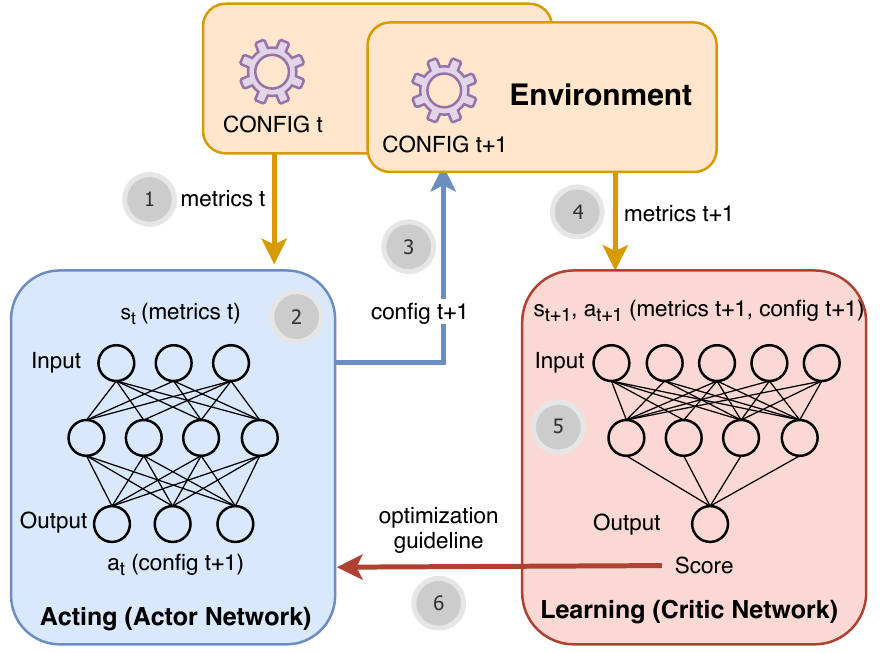}
	\centering
	\caption{A central idea of DDPG: 
	The actor network learns a policy for recommending a configuration for a given state, and the critic network on the right side evaluates the score to take an action (configuration) under a certain state (metrics). 
	This score is used as feedback to improve the actor network.}
	\label{fig:ddpg_network}
\end{figure}

\textbf{Acting procedure}:
\begin{enumerate}
	\item We retrieve the metrics of the DFS as the current state of the environment $s_t$.
	
	\item The current state $s_t$ is fed to the actor neural network, our policy, to recommend a new action $a_{t+1}$.
	
	\item The new action (configuration) $a_{t+1}$ is applied to the DFS and triggers the restart of either workloads or DFS. 
	If the static parameters involve only workloads, then only workloads need to be restarted, otherwise both the DFS and its workloads need to be restarted.

\end{enumerate}

\textbf{Learning procedure}:
\begin{enumerate}
	\item We need to fetch from memory pool a set D of transitions $(s_t, a_t, r_t, s_{t+1})$.
	
	\item The transitions are supplied to the critic network, Q function $Q_\phi (s_t, a_t$). 
	It outputs a score that estimates the benefit of taking a specific action $a_t$ at the state $s_t$.
	
	\item Based on the Bellman Equation, the critic network is optimized by minimizing the difference between estimates $Q_{\phi}(s_t, a_t)$ and target $\left(r_t+\gamma Q_{\phi_{\mathrm{targ}}}\left(s_{t+1}, \pi_{\theta_{\mathrm{targ}}}\left(s_{t+1}\right)\right)\right)$. 
		\begin{equation*}
		    \operatorname*{argmin}_\theta
			\resizebox{0.85\linewidth}{!}{%
			$
			\underset{\left(s_t, a_t, r_t, s_{t+1}\right) \sim {D}}{\mathrm{E}}\left[\left(Q_{\phi}(s_t, a_t)-\left(r_t+\gamma Q_{\phi_{\mathrm{targ}}}\left(s_{t+1}, \pi_{\theta_{\mathrm{targ}}}\left(s_{t+1}\right)\right)\right)\right)^{2}\right]
			$
			}
		\end{equation*}
	where $\theta_{\mathrm{targ}}$ stands for target network\footnote{The target network is the copy of the critic network but updated with a delay. It is used because we use and update the critic network simultaneously. By using target network, the optimization procedure is more stable.}.
	\item The actor network views the critic network ($Q_\theta$) as a fixed network, and uses it as the target to optimize its policy $\mu_\theta$,
		\begin{equation*}
		\operatorname*{argmax}_\theta  \underset{s \sim {D}}{{\mathrm E}}\left[ Q_{\phi}(s, \mu_{\theta}(s)) \right]
		\end{equation*}

\end{enumerate}

\subsubsection{Action Mapping}
\label{subsection:action_mapping}
Previous works like CAPES\cite{li2017capes} discretize continuous parameters, which could cause curse of dimensionality with a fined granularity or missing optimal configurations with a coarse granularity. 
DDPG utilizes DQN\cite{mnih2015human} and actor-critic to learn the optimal configuration from high dimensional continuous parameter space directly. However, the standard DDPG does not work with discrete parameters. 
To cope with DFS parameter tuning which contains many discrete parameters, we utilize discretization to map DDPG's action to corresponding valid values for each tuning parameter if needed. 
Therefore, the action space is normalized to $[0,1]^n$ and then inverse mapped to its actual value. 
The following equation illustrates how an action $a(i)$ is reverse mapped to its corresponding value for parameter $\lambda_i$:

\begin{equation*}
\lambda_i= 
\resizebox{0.85\linewidth}{!}{
$\begin{cases}
    a(i) \times ({\lambda_{i\_{max}} - \lambda_{i\_{min}}}) + \lambda_{i\_{min}} &,\lambda_i \text{ is continuous}\\
\lfloor a(i) \times ({\lambda_{i\_{max}} - \lambda_{i\_{min}}}) + \lambda_{i\_{min}} + 0.5\rfloor &, \lambda_i \text{ is discrete}
\end{cases}$
}
\end{equation*}
Here, $\lambda_{i\_{min}}$ and $\lambda_{i\_{min}}$ denote the maximum and minimum value of parameter $\lambda_i$ respectively.

 \subsection{Replay Buffer}
Static parameter tuning is expensive due to restarts of workloads or DFS. 
Though DDPG is an off-policy algorithm, which makes it sample efficient, to make full use of scarce tuning experiences, we adopt the replay buffer to store and learn previous experiences. 
There are two advantages of the replay buffer. 
On the one hand, we can learn from sampled experiences in different timesteps, which is more representative. 
On the other hand, batch experiences stabilize learning.

Nevertheless, learning too much from historical data might make a model overfit and not reflect reality, and vice versa. 
Therefore, in Magpie, we use a limited-sized replay buffer obeying the FIFO principle. 
Once a new transition is added and the buffer reaches its limit, the oldest data will be removed.

\section{Implementation and Evaluation}
In this section, we present the implementation and evaluation of Magpie. We first present the prototype implementation. Furthermore, we compare the performance of Magpie against a general automatic parameter tuning approach, BestConfig\cite{zhu2017bestconfig}. 
Experiments for both single and multiple performance optimizations are conducted.

\subsection{Implementation}
Our prototype system is mainly written in Python. We evaluate our system by tuning a Lustre file system (Lustre FS). 
In general, Magpie can also tune other DFSs as it is designed and implemented to be agnostic to a specific DFS. We use TF-Agent\footnote{Tensorflow Agent, \url{https://github.com/tensorflow/agents}} as our RL framework. To fetch metrics from the system, we use Telegraf\footnote{\url{https://github.com/influxdata/telegraf}} to collect metrics from the DFS and clients. InfluxDB\footnote{\url{https://github.com/influxdata/influxdb}} is the time series database used to store and query all metrics. If the tuning system has an existing metrics collection system, Magpie does not need to deploy Telegraf and InfluxDB, but can directly use the existing ones.

\begin{table}[h!]
\centering
\caption{Metrics description}
\label{tab:metrics_desc}
\begin{tabular}{p{0.25\linewidth}p{0.12\linewidth}p{0.5\linewidth}}
\toprule
\thead{Metrics} & \thead{Scope} & \thead{Description}                                                                                                              \\
\midrule
cur\_dirty\_bytes               & OSC            & Current number of bytes written and cached by this OSC.                                                                       \\  
cur\_grant\_bytes               & OSC            & Amount of space this client has reserved for writeback cache.\\
read\_rpcs\_in\_flight          & OSC            & Number of read RPCs issued, but not complete during snapshot. \\  
write\_rpcs\_in\_flight         & OSC            & Number of write RPCs issued, but not complete during the snapshot. \\  
pending\_read\_pages            & OSC            & Number of pending read pages that have been queued for I/O in the OSC.                                                             \\  
pending\_write\_pages           & OSC            & Number of pending write pages that have been queued for I/O in the OSC.                                                            \\  
cache\_hit\_ratio               & OSC    & Ratio of hit and total cache access.                                                                                 \\  
cpu\_usage\_idle                & MDS            & Amount of time that the CPU is idle.                                                                                          \\  
cpu\_usage\_iowait              & MDS            & Amount of time that the CPU is waiting for I/O operations to complete.                                                        \\  
ram\_used\_percent              & OSC\&MDS         & Used RAM percentage.                                                                                                               \\ 
\bottomrule
\end{tabular}
\end{table}
Choosing suitable metrics to represent the state of DFS is fundamental for Magpie to understand the system. All those metrics relevant to the optimization of performance indicator(s) should be included. Therefore, we consider the metrics from both DFS server and client. 
Specifically, we choose the parameters listed in~\autoref{tab:metrics_desc}. 
The column scope indicates whether it is a server or client metric. For example, the parameter in OSC\footnote{Object Storage Client (OSC) is the client service in Lustre DFS.} scope represents the status of Lustre clients. max\_rpc\_in\_flight is such a metric which describes the current number of RPC in flight issued by the Lustre client. This metric helps Magpie to observe the status of client requests. Apart from client metrics, we use memory performance indicators to observe the load status of MDSs\footnote{MetaData Service (MDS) is the metadata service in Lustre DFS.} and cpu usage indicators for both servers and clients (MDSs and OSTs).

We tune the following two static parameters:
\begin{enumerate}
    \item stripe\_count: The number of OSTs is written across.
    \item stripe\_size:  The chunk size to stripe files.
\end{enumerate}

\subsection{System and Experiment Setup}

\subsubsection{DFS}
We tune a Lustre FS in our evaluation. Six nodes are dedicated to run Lustre FS and three nodes for Lustre clients. The Lustre version we use for server and client are both release 2.12.7. Each node has the same physical hardware and OS, i.e., every node is equipped with 16 GB RAM, 3x 1TB Hitachi HDD, and is connected via a single 1 GB switch.

\subsubsection{Workload}
To better simulate real workloads, we use Filebench\footnote{\url{https://github.com/filebench/filebench}}. Filebench is a widely-used toolkit which can simulate various applications' I/O behaviors by using its Workload Model Language (WML). As shown in~\autoref{tab:workload_description}, we use a selection of predefined Filebench workloads. During the training session, each workload is running for two minutes and the final configurations are evaluated for 30 minutes with three runs.

\begin{table}[h]
\caption{Workloads description}
\label{tab:workload_description}
\begin{tabular}{p{0.25\columnwidth}p{0.65\columnwidth}}
\toprule
\thead{Workload}          & \thead{Description} \\ \midrule
File Server       & Emulates simple file-server I/O activities. This workload performs a sequence of creates, deletes, appends, reads, writes and attribute operations.                                                                                                  \\
Video Server      & Emulates a video server, which provide two file sets, active and inactive videos, for access.\\
Sequential Write & Sequential write of 5 files using multiple threads.\\ 
Sequential Read  & Sequential read of 5 files using multiple threads. \\
Random RW & Two threads working on the same large file. One thread does random reads, and another does random writes.
\\ \bottomrule
\end{tabular}
\end{table}

\begin{figure}[hb!]
    \centering
    \includegraphics[width=\linewidth]{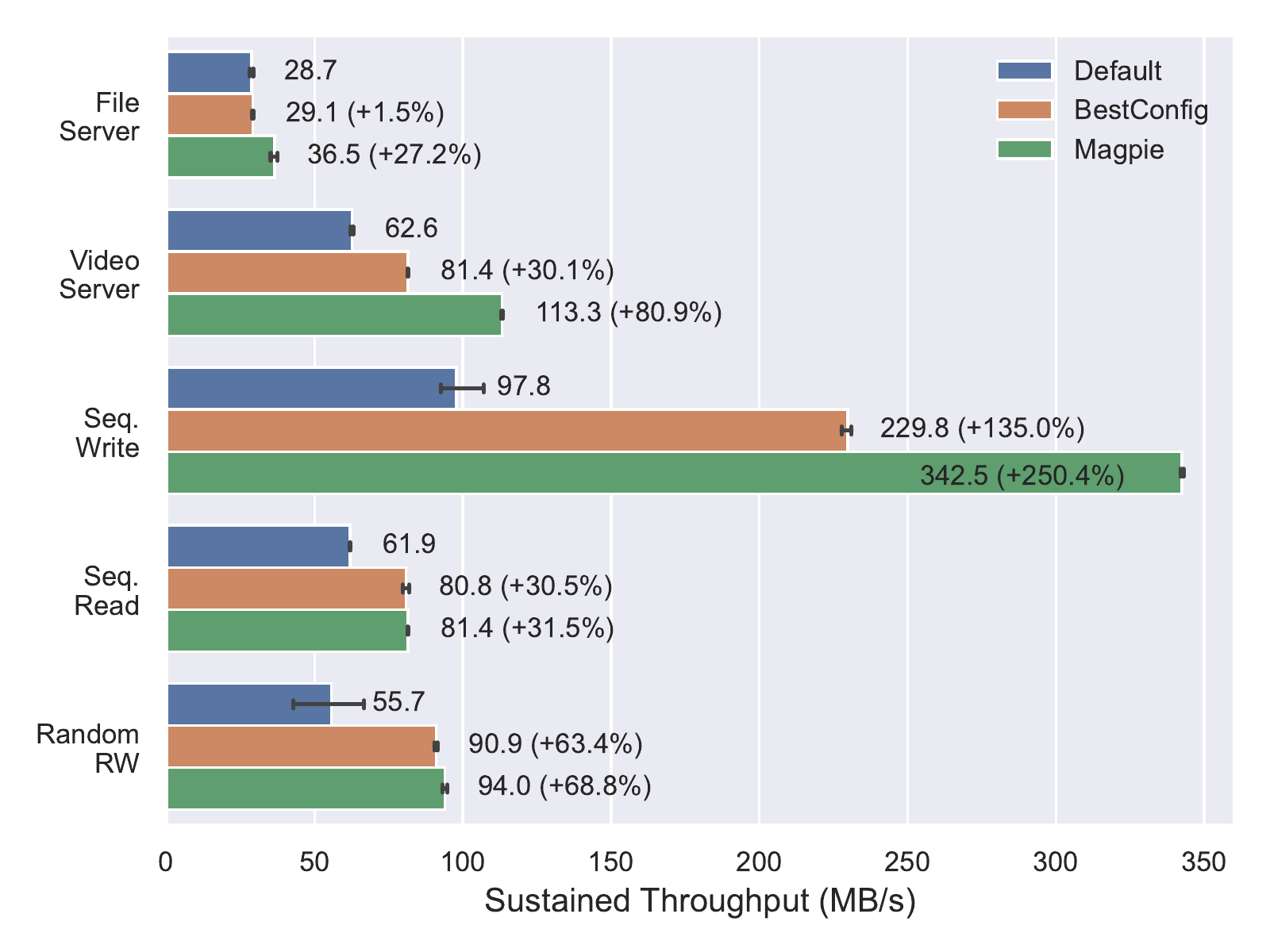}
    \caption{Tuning throughput under different workloads. Magpie achieves noticeable performance gains and outperforms BestConfig on all workloads.}
    \label{fig:single_optimization_tuning_performance_comparison}
\end{figure}
\subsubsection{Magpie}
There is only one node dedicated to train the Magpie model. This node is equipped with a NVIDIA Quadro RTX 5000.
On this node, we run an InfluxDB server and our RL model. The version of InfluxDB is 2.1.1. On all Lustre servers and client nodes, Telegraf 1.19.2 is installed to monitor and collect performance indicators.

\begin{figure*}[ht!]
    \centering
    \subfloat[File Server]{{\includegraphics[width=.48\textwidth]{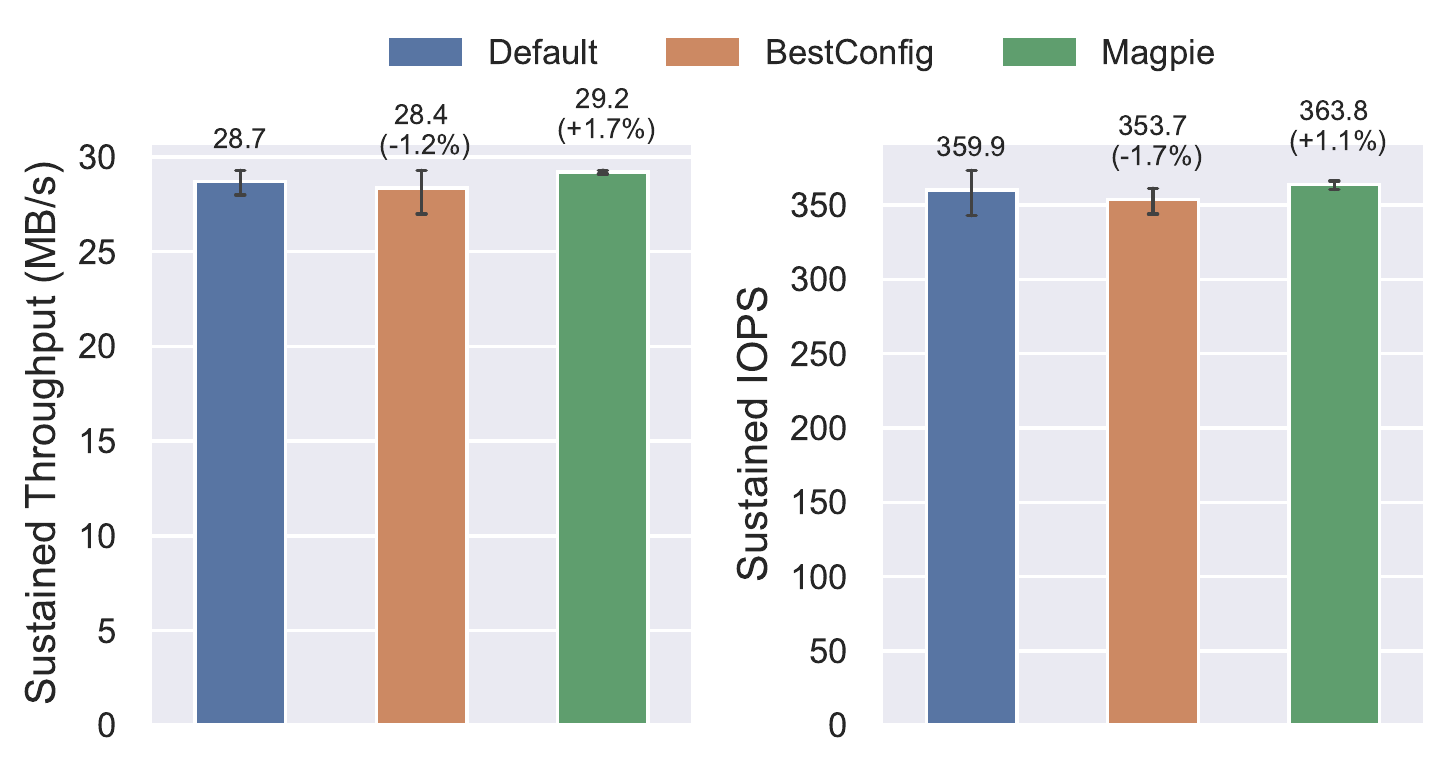} }}\hfill
    \subfloat[Video Server]{{\includegraphics[width=.48\textwidth]{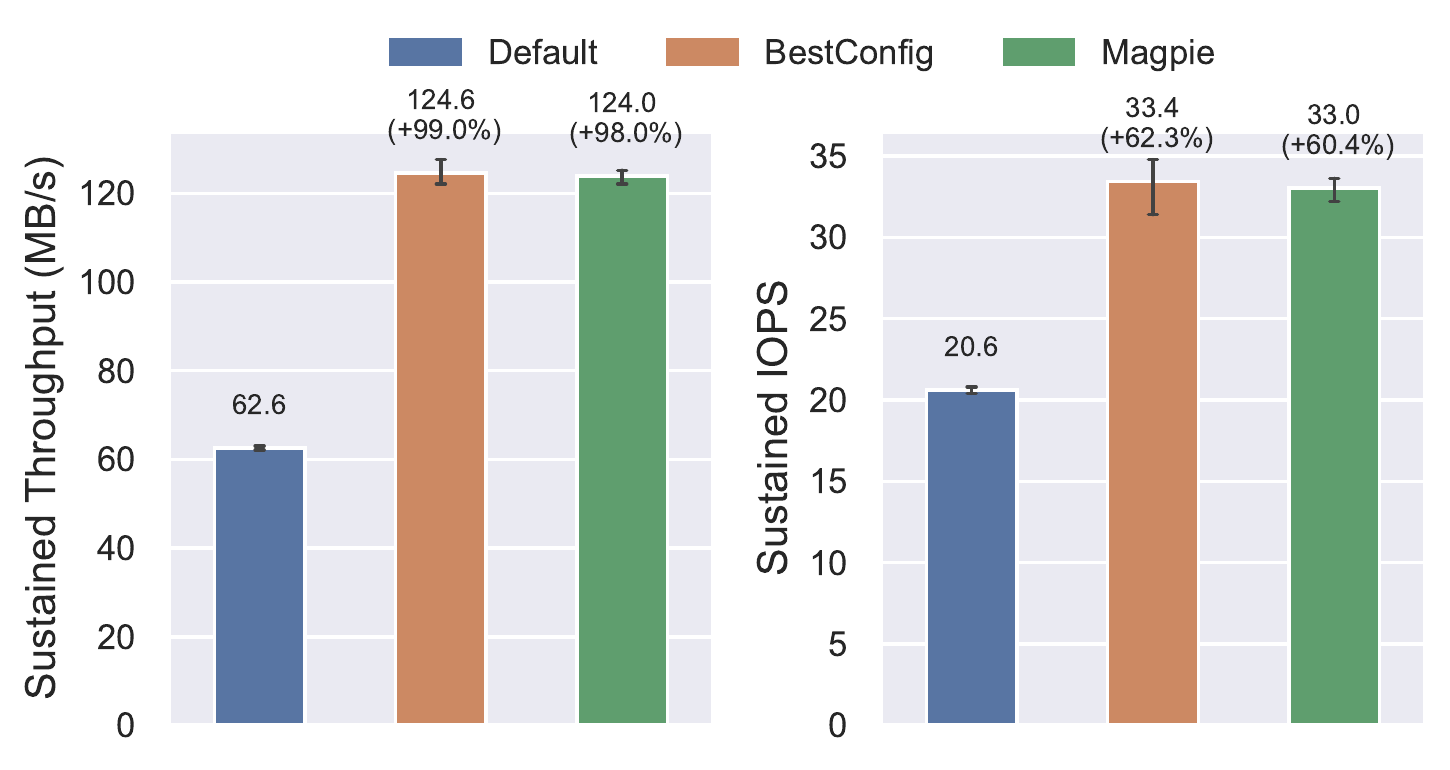} }}\par
    \subfloat[Sequential Write]{{\includegraphics[width=.48\textwidth]{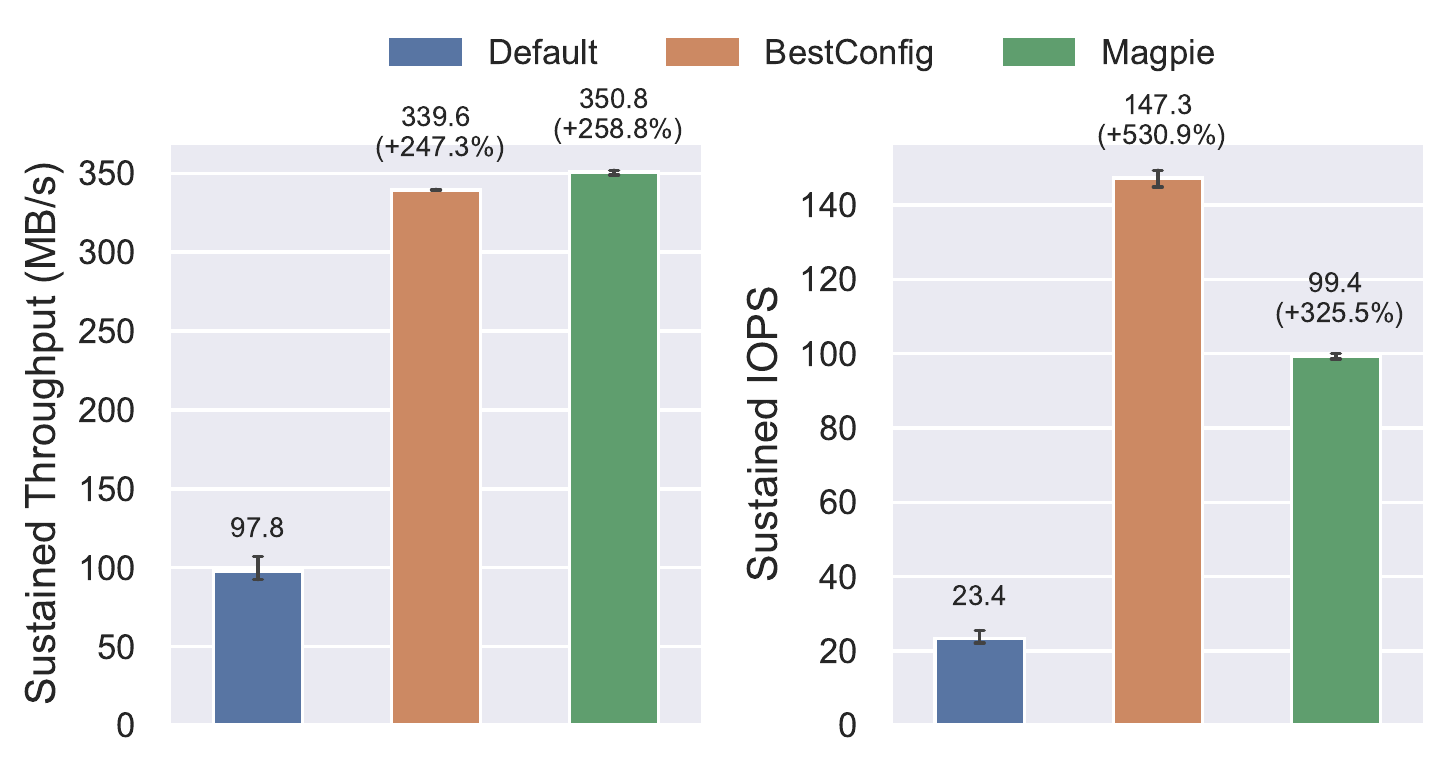} }}\hfill
    \subfloat[Sequential Read]{{\includegraphics[width=.48\textwidth]{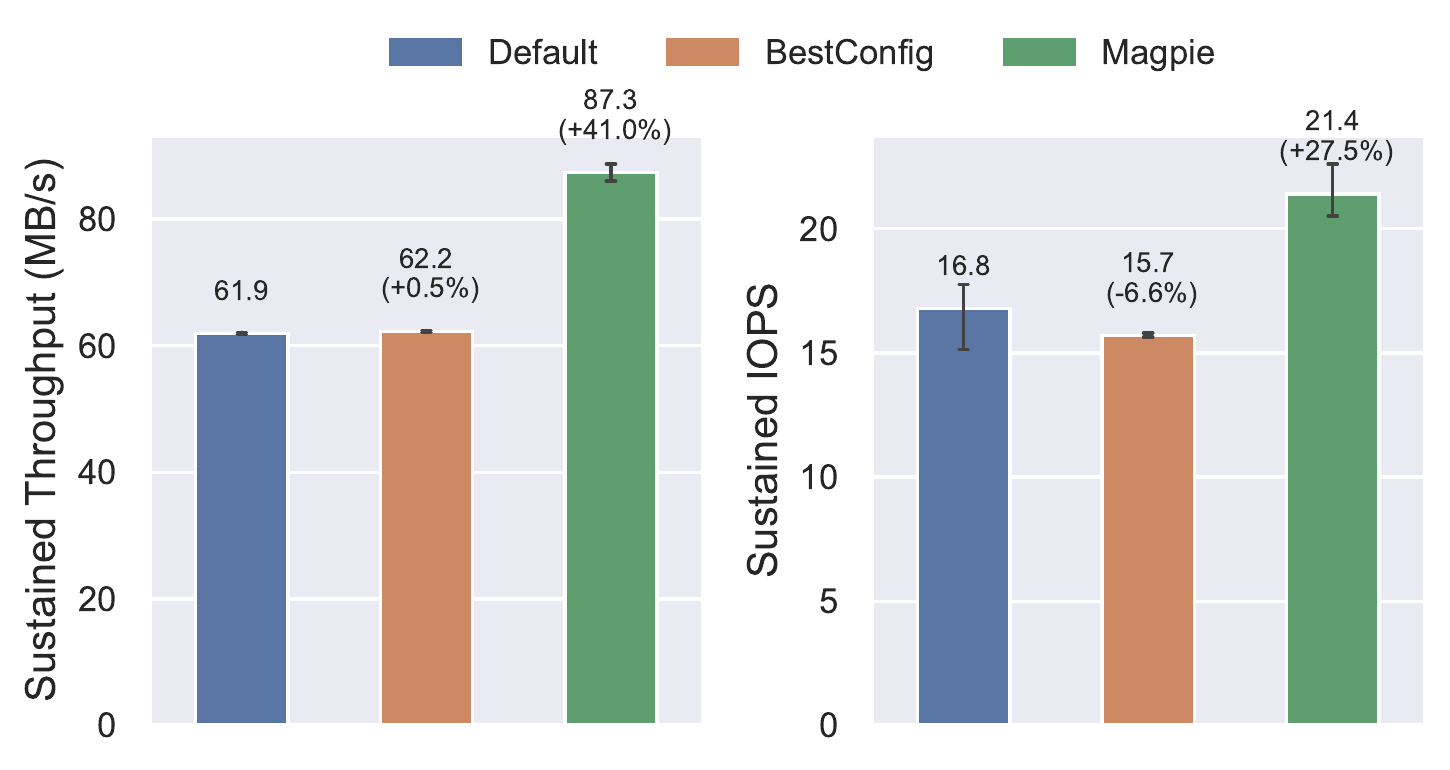}}}
    \par
    \subfloat[Random RW]{{\includegraphics[width=.48\textwidth]{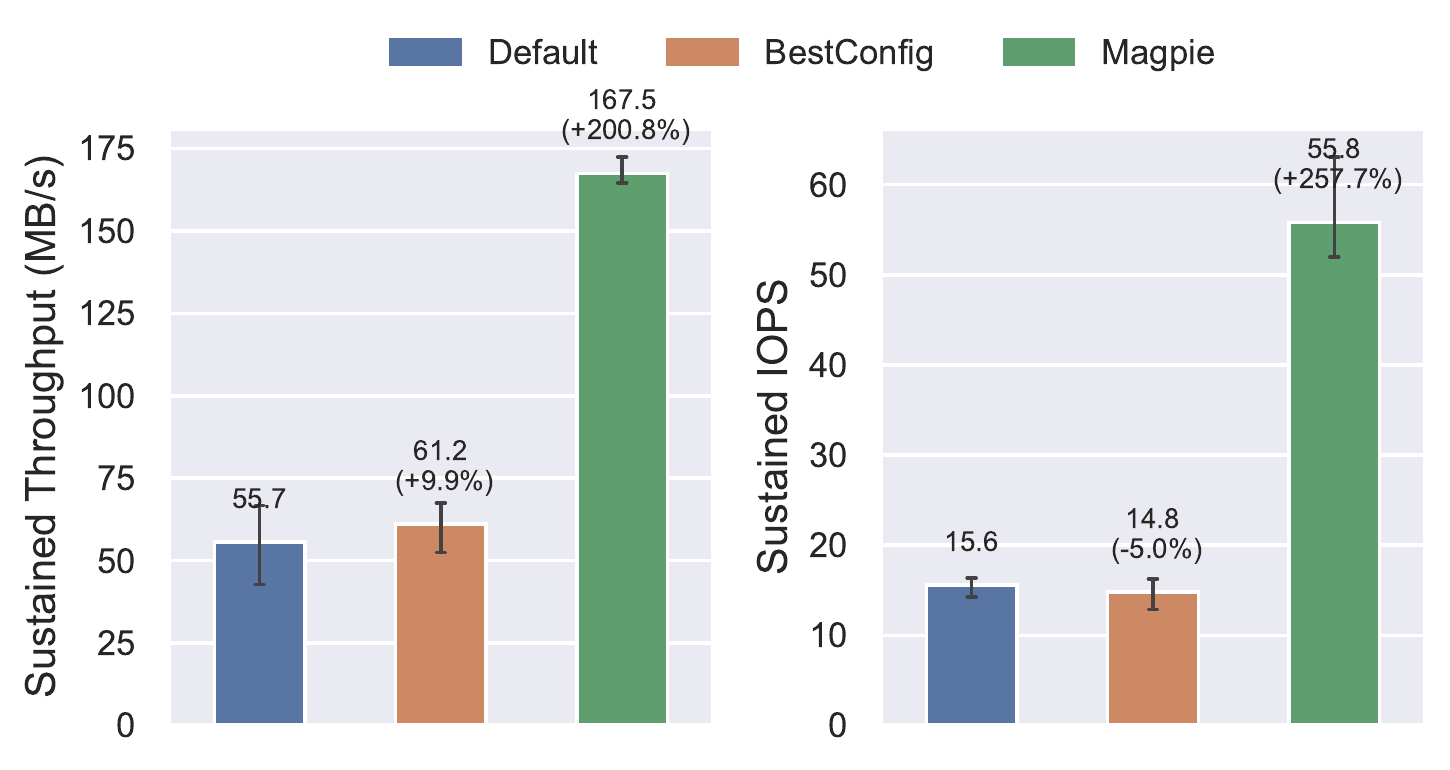} }}
        
    \caption{Optimizing performance of throughput and IOPS in parallel. Magpie achieves noticeable performance gains for both metrics on most workloads.}
    \label{fig:double_optimization_result}
\end{figure*}

\subsection{Single Performance Indicator Optimization}
\label{subsub:single_optimization}
In this experiment, we evaluate the tuning performance under five FileBench workloads. \emph{File Server} and \emph{Video Server} are commonly used with DFS \cite{duan2015high}, and the other three workloads are regular IO operations in DFS. As there is no existing automatic parameter tuning system that can directly tune static parameters, we employ a general parameter tuning system, BestConfig\cite{zhu2017bestconfig}, as our baseline. In each setup, both methods take 30 tuning actions and are evaluated three times to take the average as the final performance.

As shown in~\autoref{fig:single_optimization_tuning_performance_comparison}, Magpie outperforms BestConfig for all workloads, especially in the Sequential Write, where Magpie achieves a 250.4\% performance gain, which is almost twice that of BestConfig.

\subsection{Multiple Performance Indicators Optimization}
\label{subsub:multip_optimization}
The optimization of multiple performance indicators requires the tuning system to not only optimize a single performance metric but also improve others simultaneously. In this experiment, we evaluate Magpie's performance to optimize Throughput and Input/Output operations Per Second (IOPS) at the same time. The metric throughput quantifies the rate of successfully delivered messages while IOPS measures the number of input and output operations per second. Operations like Online Transaction Process (OLTP) favor IOPSs, whereas data-intensive type operations prefer throughput. However, a workload mixed with OTLP and data-intensive operations demands both high throughput and IOPSs. 

We use Magpie to tune five workloads to evaluate its capabilities of multiple objective optimization. The experiment results are shown in~\autoref{fig:double_optimization_result}. Magpie achieves noticeable performance gains in all workloads, except File Server. Compared to BestConfig performance, Magpie achieves similar performance gains in File Server and Video Server workload. In workload Sequential Write, Magpie achieves more gain than BestConfig on throughput while less on IOPS. Magpie outperforms BestConfig in the other two workloads.

\begin{figure}[h!]
    \centering
    \includegraphics[width=.48\textwidth]{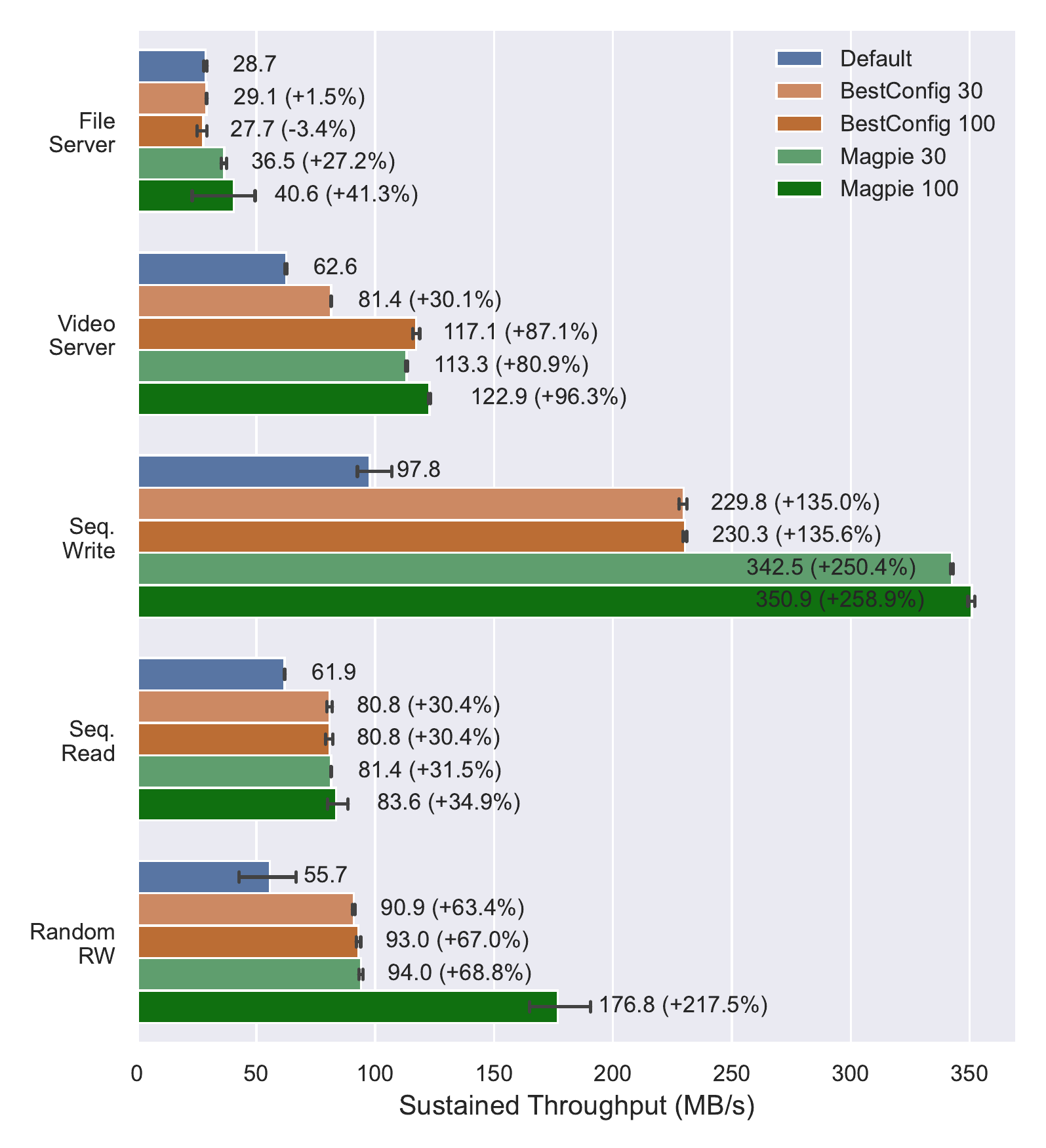}
    \caption{Performance gains after tuned with a different number of training steps. 30 and 100 stand for maximal tuning steps. More training steps lead to higher performance gains in Magpie.}
    \label{fig:soo_tuning_step_perf_comp}
\end{figure}
\subsection{Tuning Efficiency and Flexibility}
We conduct two experiments to illustrate the tuning efficiency and flexibility of Magpie. In the first experiment, we tune the Lustre file system to optimize throughput with Magpie and BestConfig for 70 additional tuning steps, i.e., 100 tuning steps in total. Both BestConfig and Magpie utilize previous tuning experience. For example, Magpie with 100 tuning steps (Magpie 100) makes use of the tuning experience from Magpie with 30 tuning steps (Magpie 30), i.e., it starts tuning from iteration 31. As presented in~\autoref{fig:soo_tuning_step_perf_comp}, Magpie can achieve sufficient performance gains with 30 steps in all workloads, though with more training steps it always attains higher performance gains, especially for the workload Random RW. On the contrary, BestConfig 100 does not gain much performance improvement with more training steps, except for the Video Server workload. 
Also, BestConfig 100 performs worse than BestConfig 30 for the File Server workload. This might be due to the high variance of this workload, which hinders appropriate learning in BestConfig.

The second experiment is to investigate the influence of the number of tuning steps on BestConfig and Magpie. We run both approaches progressively on the Video Server workload and track its tuning performance changes. Specifically, we tune 10 more steps on top of the previous tuning experience in each new run. As presented in~\autoref{fig:training_speed}, apart from the progressive tuning results of Progressive BestConfig and Progressive Magpie, BestConfig 30 and BestConfig 100 are also depicted as BestConfig 30 \& 100 for comparison. Progressive Magpie implicitly includes Magpie 30 and Magpie 100 because Magpie internally is a progressive approach.
Empirical results show that Magpie makes significant performance gains with 10 tuning steps and then it utilizes the additional tuning steps for fine-tuning. The plateau in steps 10-50 and steps 60-100 is because Magpie does not find better parameter values, hence it recommends the best it has seen so far. 
On the contrary, BestConfig performs poorly at small-step progressive tuning (Progressive BestConfig) but performs well at big-step progressive tuning (BestConfig 30 \& 100). We assume this is because BestConfig relies on initial sampling to approach optimal configurations. If it does not sample near the optimal configurations in the first round, it will be trapped with a sub-optimal solution. Consequently, small-step lowers the chance for BestConfig to find the optimal configurations. 
Additionally, for the slight performance drop in training steps 10-20, we assume the fluctuation of throughput misleads BestConfig and accordingly favors the selection of slightly worse configurations.

\begin{figure}[h!]
    \centering
    \includegraphics[width=.48\textwidth]{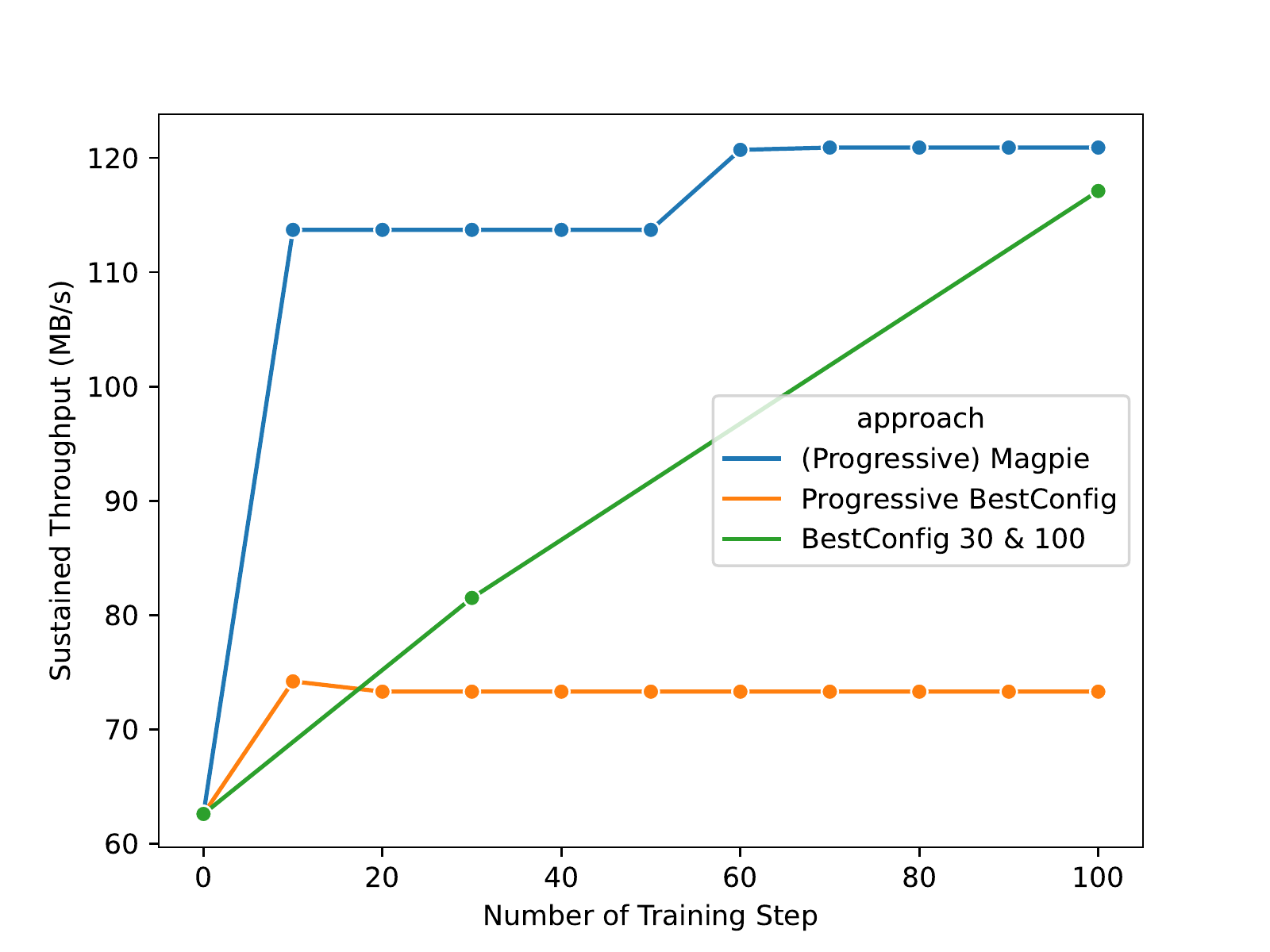}
    \caption{Tuning results with the different number of tuning steps for the Video Server workload. Magpie achieves noticeable performance gains with 10 tuning steps and then uses additional tuning steps for parameter fine-tuning.}
    \label{fig:training_speed}
\end{figure}

As shown in these two experiments, Magpie can optimize not only noticeably with a small number of tuning steps, but also qualifies for continuous tuning on top of previous experiments when more tuning cost is provided. As a result, users can flexibly decide during the training if they are satisfied with the achieved performance improvements using low tuning cost or wait longer to get the optimal configuration. Furthermore, users can still resume tuning when they want to reach an even higher performance at a later point in time.

\subsection{Tuning Cost}
The costs of employing Magpie for tuning static parameters can be unraveled as follows:

\begin{enumerate}[label=\alph*.]
    \item Software (Optional).
    The additional software we need is the performance collecting agent and collector, i.e., Telegraf and InfluxDB. As Magpie is implemented agnostic to a specific system, if there is such a collecting agent present, Magpie can utilize existing software to fetch metrics. 
    
    \item Hardware.
    To use Magpie, we need one additional server to run Magpie as well as InfluxDB for collecting performance indicators. However, oftentimes users are utilizing a central metrics collecting system anyways. In that case, we only need to have a server to deploy Magpie.
    
    \item Downtime.
    Each time when we change values of static parameters, the workload or DFS needs to be restarted. In our setting, it takes 12 to 20 seconds to restart the workload and about 30 seconds to restart the DFS.
    
    \item Tuning Operation. The time cost of all operations is shown in~\autoref{tab:execution_time}.
\end{enumerate}

\begin{table}[ht!]
	\ra{1.3}
	\caption{Technical measurements of the Magpie evaluation} 
	\label{tab:execution_time}
	\centering
	\begin{tabular}{@{}p{0.25\linewidth}p{0.2\linewidth}p{0.4\linewidth}@{}}
		\toprule
		\thead{Name} & \thead{Time} & \thead{Description} \\ 
		\midrule
		Action step time  	& 	3.5s	& Runtime of taking an action and retrieving the performance indicators. \\
		Model update time 	& 	0.72s& Runtime of training the agent for the given number of training steps after each action step. \\
		One iteration time 	& 	4.8s & Runtime of taking one iteration (includes action step time, training step time and other operation time). \\
	   \bottomrule
  \end{tabular}
\end{table}

\section{Related Work}
Tuning configuration parameters for DFSs is a demanding task. The optimal configuration depends on systems, workloads, and even the version of the software. Many approaches have been proposed in order to find the best parameter setting. 
These approaches can be broadly separated into two categories: search-based and model-based methods. We will present the related works in these categories in the following two subsections.
These works focus on dynamic parameters, and cannot be applied to static DFS parameters directly, and to the best of our knowledge, there is no previous work targeting the tuning of static parameters in DFSs.

\subsection{Search-Based Methods}
Search-based approaches employ no information from the DFS or workloads, instead, concentrate on searching the parameter space directly. 
Typically they work by making clever assumptions about the parameter space and sampling a small set of possible candidates. The gained knowledge is used to make a more informed decision in another round of sampling the parameter space. This is then usually repeated in an iterative process.

BestConfig~\cite{zhu2017bestconfig} is such a search-based approach. It proposes a Divide and Diverge Sampling method to divide each parameter into a certain number of intervals and then diverge (choose) the sample points by representing each interval of each parameter exactly once. They additionally employ another step to optimize the sampling performance, this so-called Recursive Bound and Search method assumes a higher possibility of finding better performance around the point with the best performance from the initial sample points. BestConfig then searches recursively in a bounded space around the best-performing point.

One group of search-based methods is based on Bayesian Optimization (BO) approaches, which views the objective function as a stochastic process.
They are suitable for the evaluation of expensive-to-evaluate situations and can tolerate stochastic noise in function evaluations~\cite{cao2018towards,frazier2018tutorial}. 
For example, BOAT~\cite{dalibard2017boat} utilizes structured BO to tune general systems. It utilizes system developers' context information to build a probabilistic model to predict the behavior of the system and then make informed decisions on parameter configurations to be evaluated.
Wu et al. use Gaussian processes to find the relationship between a machine learning model's performance and its hyper-parameters~\cite{wu2019hyperparameter}. They frame the hyper parameter tuning problem as an optimization problem, and then use Bayesian optimization to find the optimal parameters.
Berkenkamp et al. use a safe Bayesian optimization algorithm (SAFEOPT) for automatic parameter tuning in robotics~\cite{berkenkamp2021bayesian}. Their approach can work with multiple safety constraints and aims to optimize parameters while ensuring performance above a certain threshold.

Another group is Evolutionary Algorithms (EA). They are inspired by natural selection and can quickly adapt to changes. They are widely used in many complex problems due to their robustness, flexibility, and conceptual simplicity~\cite{vikhar2016evolutionary}. Saboori et al. employ Covariance Matrix Adaptation (CMA), an EA, to auto-tune distributed systems~\cite{saboori2008autotuning}. 
Several researchers have utilized Evolutionary Algorithms to optimize the hyper-parameters of Machine Learning and Deep Learning models~\cite{lorena2008evolutionary,young2015optimizing,zhang2021convolutional}.

Different from search-based methods, Magpie considers both server and client metrics to interpret the parameters' influence on performance. Correspondingly, it demands a limited amount of samples for tuning while search-based methods oftentimes need numerous sampling points to find good configurations.

\subsection{Model-Based Methods}
Model-based methods model the relation between the configuration parameters, workloads, and the performance of the DFS.
In Supervised Machine Learning, the task is to learn an input-output mapping function from sampled data points. With enough representative data points and a strong enough variable relevance, models learned through Supervised Machine Learning algorithms can often generalize the relationship with a good performance.

For example, OtterTune utilizes machine learning to learn a model from historical data which recommends a configuration~\cite{van2017automatic}. Yigitbasi et al. propose a machine learning-based method to build an end-to-end tuning flow for MapReduce~\cite{yigitbasi2013towards}. Chen et al. use two phases, prediction and evaluation, to improve the performance of MapReduce jobs, such that optimal configuration can be found via repetitively invoking predictors~\cite{chen2015machine}. Enel uses a graph model to predict the runtime of tasks in data flow jobs and tune the scale-out of tasks to meet the runtime target requirements~\cite{scheinert2021enel}. Nevertheless, training data is not always easy to acquire, and unrepresentative data often leads to poor performance of supervised learning models.

Reinforcement Learning (RL) is a branch of machine learning, which adapts ideas from psychological learning theory and solves optimization problems by trial and error~\cite{sutton2018reinforcement}. Due to its flexibility, it can well suit the changeable workload requirements in storage usage.
CAPES~\cite{li2017capes} is a simple neural network-based approach for distributed storage performance tuning. It takes periodic measurements of target storage systems and suggests a change to the parameter value. CDBTune~\cite{zhang2019end} and $\text{CDBTune}^+$~\cite{zhang2021hbox} are auto-tuning methods for cloud databases performance optimization using Deep RL~. They utilize Deep Deterministic Policy Gradient (DDPG) to tune cloud databases in high-dimensional continuous space.
QTune\cite{li2019qtune} is another approach which uses DDPG to tune database configurations. It features SQL queries by considering rich features (e.g., query type, tables) within the queries and then selects the optimal configuration according to the DRL model. 

Unlike other model-based approaches, Magpie requires no historical data in advance and utilizes system metrics to understand the relation between parameter values and system performance. CDBTune, $\text{CDBTune}^+$ and QTune are the methods most related to Magpie. However, they mainly aim to tune knobs of databases, to which tuning overhead is not critical, whereas Magpie works on DFS parameters with significant tuning costs. Furthermore, QTune makes use of specific database properties like SQL query type, while Magpie employs internal and external metrics of DFS to speed the tuning process and decrease the tuning cost. 
Another closely related work is CAPES. It tunes dynamic parameters of DFSs but can only change the parameter value within a step size per tuning action, which is time-consuming and disqualifying in static parameter tuning. Therefore, those related works could not be applied directly to the static parameters tuning of DFSs.

\section{Conclusion}
This paper presented Magpie, an automatic tuning system to tune static parameters of distributed file systems.
It incorporates server-side and client-side metrics to better discover the relation between parameters and performance. Furthermore, it can automatically tune static parameters for DFSs using Deep Reinforcement Learning. Lastly, it also supports optimizing multiple performance indicators simultaneously.

In our experiments, we evaluated Magpie in two different settings, first optimization of a single performance indicator and then also multiple performance indicators. We compared our approach with a general parameter tuning system, BestConfig. The tuning results showed that our method not only achieves noticeable performance gains after tuning, but also outperforms BestConfig. Specifically, in single performance indicator optimization, our approach on average achieved 91.8\% throughput gains against the default configuration and 39.7\% compared to the baseline. In multiple performance indicator optimization, our approach on average reached 119.4\% throughput gains and 272.8\% IOPS gains compared to the default configuration, while it yielded 48.3\% more throughput and 156.8\% more IOPS compared to the baseline method.

In the future, we want to further investigate boosting the tuning of static parameters by employing insights learned from other workloads.

\balance

\bibliographystyle{IEEEtran}
\bibliography{references}

\begin{thebibliography}{10}
\providecommand{\url}[1]{#1}
\csname url@samestyle\endcsname
\providecommand{\newblock}{\relax}
\providecommand{\bibinfo}[2]{#2}
\providecommand{\BIBentrySTDinterwordspacing}{\spaceskip=0pt\relax}
\providecommand{\BIBentryALTinterwordstretchfactor}{4}
\providecommand{\BIBentryALTinterwordspacing}{\spaceskip=\fontdimen2\font plus
\BIBentryALTinterwordstretchfactor\fontdimen3\font minus
  \fontdimen4\font\relax}
\providecommand{\BIBforeignlanguage}[2]{{%
\expandafter\ifx\csname l@#1\endcsname\relax
\typeout{** WARNING: IEEEtran.bst: No hyphenation pattern has been}%
\typeout{** loaded for the language `#1'. Using the pattern for}%
\typeout{** the default language instead.}%
\else
\language=\csname l@#1\endcsname
\fi
#2}}
\providecommand{\BIBdecl}{\relax}
\BIBdecl

\bibitem{braam2019lustre}
P.~Braam, ``The lustre storage architecture,'' \emph{arXiv preprint
  arXiv:1903.01955}, 2019.

\bibitem{weil2007ceph}
S.~A. Weil, ``Ceph: reliable, scalable, and high-performance distributed
  storage,'' Ph.D. dissertation, University of California, Santa Cruz, 2007.

\bibitem{zadok2015parametric}
E.~Zadok, A.~Arora, Z.~Cao, A.~Chaganti, A.~Chaudhary, and S.~Mandal,
  ``Parametric optimization of storage systems,'' in \emph{7th USENIX Workshop
  on Hot Topics in Storage and File Systems (HotStorage 15)}, 2015.

\bibitem{cao2018towards}
Z.~Cao, V.~Tarasov, S.~Tiwari, and E.~Zadok, ``Towards better understanding of
  black-box {Auto-Tuning}: A comparative analysis for storage systems,'' in
  \emph{2018 USENIX Annual Technical Conference (USENIX ATC 18)}.\hskip 1em
  plus 0.5em minus 0.4em\relax USENIX Association, Jul. 2018, pp. 893--907.

\bibitem{cao2020carver}
Z.~Cao, G.~Kuenning, and E.~Zadok, ``Carver: Finding important parameters for
  storage system tuning,'' in \emph{18th USENIX Conference on File and Storage
  Technologies (FAST 20)}, 2020, pp. 43--57.

\bibitem{lyu2020sapphire}
W.~Lyu, Y.~Lu, J.~Shu, and W.~Zhao, ``Sapphire: Automatic configuration
  recommendation for distributed storage systems,'' \emph{arXiv preprint
  arXiv:2007.03220}, 2020.

\bibitem{herodotou_survey_2021}
H.~Herodotou, Y.~Chen, and J.~Lu, ``\BIBforeignlanguage{en}{A {Survey} on
  {Automatic} {Parameter} {Tuning} for {Big} {Data} {Processing} {Systems}},''
  \emph{\BIBforeignlanguage{en}{ACM Computing Surveys}}, vol.~53, no.~2, pp.
  1--37, Mar. 2021.

\bibitem{li2017capes}
Y.~Li, K.~Chang, O.~Bel, E.~L. Miller, and D.~D. Long, ``{CAPES: unsupervised
  storage performance tuning using neural network-based deep reinforcement
  learning},'' in \emph{Proceedings of the International Conference for High
  Performance Computing, Networking, Storage and Analysis}, 2017, pp. 1--14.

\bibitem{zhang2019end}
J.~Zhang, Y.~Liu, K.~Zhou, G.~Li, Z.~Xiao, B.~Cheng, J.~Xing, Y.~Wang,
  T.~Cheng, L.~Liu \emph{et~al.}, ``An end-to-end automatic cloud database
  tuning system using deep reinforcement learning,'' in \emph{Proceedings of
  the 2019 International Conference on Management of Data}, 2019, pp. 415--432.

\bibitem{cao2019practical}
Z.~Cao, ``A practical, real-time auto-tuning framework for storage systems,''
  Ph.D. dissertation, State University of New York at Stony Brook, 2019.

\bibitem{scheinert2021enel}
D.~Scheinert, H.~Zhu, L.~Thamsen, M.~K. Geldenhuys, J.~Will, A.~Acker, and
  O.~Kao, ``Enel: Context-aware dynamic scaling of distributed dataflow jobs
  using graph propagation,'' in \emph{2021 IEEE International Performance,
  Computing, and Communications Conference (IPCCC)}.\hskip 1em plus 0.5em minus
  0.4em\relax IEEE, 2021, pp. 1--8.

\bibitem{cheng2021aioc2}
W.~Cheng, S.~Deng, L.~Zeng, Y.~Wang, and A.~Brinkmann, ``{$\text{AIOC}^2$: A
  deep Q-learning approach to autonomic I/O congestion control in Lustre},''
  \emph{Parallel Computing}, vol. 108, p. 102855, 2021.

\bibitem{huang2019automatic}
C.~Huang, B.~Yuan, Y.~Li, and X.~Yao, ``Automatic parameter tuning using
  bayesian optimization method,'' in \emph{2019 IEEE Congress on Evolutionary
  Computation (CEC)}.\hskip 1em plus 0.5em minus 0.4em\relax IEEE, 2019, pp.
  2090--2097.

\bibitem{young2015optimizing}
S.~R. Young, D.~C. Rose, T.~P. Karnowski, S.-H. Lim, and R.~M. Patton,
  ``Optimizing deep learning hyper-parameters through an evolutionary
  algorithm,'' in \emph{Proceedings of the workshop on machine learning in
  high-performance computing environments}, 2015, pp. 1--5.

\bibitem{chen2015machine}
C.-O. Chen, Y.-Q. Zhuo, C.-C. Yeh, C.-M. Lin, and S.-W. Liao, ``Machine
  learning-based configuration parameter tuning on hadoop system,'' in
  \emph{2015 IEEE International Congress on Big Data}.\hskip 1em plus 0.5em
  minus 0.4em\relax IEEE, 2015, pp. 386--392.

\bibitem{lorena2008evolutionary}
A.~C. Lorena and A.~C. De~Carvalho, ``Evolutionary tuning of svm parameter
  values in multiclass problems,'' \emph{Neurocomputing}, vol.~71, no. 16-18,
  pp. 3326--3334, 2008.

\bibitem{yigitbasi2013towards}
N.~Yigitbasi, T.~L. Willke, G.~Liao, and D.~Epema, ``Towards machine
  learning-based auto-tuning of mapreduce,'' in \emph{2013 IEEE 21st
  International Symposium on Modelling, Analysis and Simulation of Computer and
  Telecommunication Systems}.\hskip 1em plus 0.5em minus 0.4em\relax IEEE,
  2013, pp. 11--20.

\bibitem{zhang2021convolutional}
M.~Zhang, H.~Li, S.~Pan, J.~Lyu, S.~Ling, and S.~Su, ``Convolutional neural
  networks-based lung nodule classification: A surrogate-assisted evolutionary
  algorithm for hyperparameter optimization,'' \emph{IEEE Transactions on
  Evolutionary Computation}, vol.~25, no.~5, pp. 869--882, 2021.

\bibitem{berkenkamp2021bayesian}
F.~Berkenkamp, A.~Krause, and A.~P. Schoellig, ``Bayesian optimization with
  safety constraints: safe and automatic parameter tuning in robotics,''
  \emph{Machine Learning}, pp. 1--35, 2021.

\bibitem{wu2019hyperparameter}
J.~Wu, X.-Y. Chen, H.~Zhang, L.-D. Xiong, H.~Lei, and S.-H. Deng,
  ``Hyperparameter optimization for machine learning models based on bayesian
  optimization,'' \emph{Journal of Electronic Science and Technology}, vol.~17,
  no.~1, pp. 26--40, 2019.

\bibitem{dalibard2017boat}
V.~Dalibard, M.~Schaarschmidt, and E.~Yoneki, ``Boat: Building auto-tuners with
  structured bayesian optimization,'' in \emph{Proceedings of the 26th
  International Conference on World Wide Web}, 2017, pp. 479--488.

\bibitem{alipourfard2017cherrypick}
O.~Alipourfard, H.~H. Liu, J.~Chen, S.~Venkataraman, M.~Yu, and M.~Zhang,
  ``Cherrypick: Adaptively unearthing the best cloud configurations for big
  data analytics,'' in \emph{14th USENIX Symposium on Networked Systems Design
  and Implementation USENIX 17)}, 2017, pp. 469--482.

\bibitem{saboori2008autotuning}
A.~Saboori, G.~Jiang, and H.~Chen, ``Autotuning configurations in distributed
  systems for performance improvements using evolutionary strategies,'' in
  \emph{2008 The 28th International Conference on Distributed Computing
  Systems}.\hskip 1em plus 0.5em minus 0.4em\relax IEEE, 2008, pp. 769--776.

\bibitem{jomaa2019hyp}
H.~S. Jomaa, J.~Grabocka, and L.~Schmidt-Thieme, ``Hyp-rl: Hyperparameter
  optimization by reinforcement learning,'' \emph{arXiv preprint
  arXiv:1906.11527}, 2019.

\bibitem{neary2018automatic}
P.~Neary, ``Automatic hyperparameter tuning in deep convolutional neural
  networks using asynchronous reinforcement learning,'' in \emph{2018 IEEE
  international conference on cognitive computing (ICCC)}.\hskip 1em plus 0.5em
  minus 0.4em\relax IEEE, 2018, pp. 73--77.

\bibitem{van2017automatic}
D.~Van~Aken, A.~Pavlo, G.~J. Gordon, and B.~Zhang, ``Automatic database
  management system tuning through large-scale machine learning,'' in
  \emph{Proceedings of the 2017 ACM International Conference on Management of
  Data}, 2017, pp. 1009--1024.

\bibitem{zhu2017bestconfig}
Y.~Zhu, J.~Liu, M.~Guo, Y.~Bao, W.~Ma, Z.~Liu, K.~Song, and Y.~Yang,
  ``Bestconfig: tapping the performance potential of systems via automatic
  configuration tuning,'' in \emph{Proceedings of the 2017 Symposium on Cloud
  Computing}, 2017, pp. 338--350.

\bibitem{zhang2021hbox}
J.~Zhang, K.~Zhou, G.~Li, Y.~Liu, M.~Xie, B.~Cheng, and J.~Xing, ``{CDBTune+}:
  An efficient deep reinforcement learning-based automatic cloud database
  tuning system,'' \emph{The VLDB Journal}, vol.~30, no.~6, pp. 959--987, 2021.

\bibitem{li2019qtune}
G.~Li, X.~Zhou, S.~Li, and B.~Gao, ``Qtune: A query-aware database tuning
  system with deep reinforcement learning,'' \emph{Proceedings of the VLDB
  Endowment}, vol.~12, no.~12, pp. 2118--2130, 2019.

\bibitem{van2014multi}
K.~Van~Moffaert and A.~Now{\'e}, ``Multi-objective reinforcement learning using
  sets of pareto dominating policies,'' \emph{The Journal of Machine Learning
  Research}, vol.~15, no.~1, pp. 3483--3512, 2014.

\bibitem{mnih2015human}
V.~Mnih, K.~Kavukcuoglu, D.~Silver, A.~A. Rusu, J.~Veness, M.~G. Bellemare,
  A.~Graves, M.~Riedmiller, A.~K. Fidjeland, G.~Ostrovski \emph{et~al.},
  ``Human-level control through deep reinforcement learning,'' \emph{nature},
  vol. 518, no. 7540, pp. 529--533, 2015.

\bibitem{duan2015high}
H.~Duan, W.~Zhan, G.~Min, H.~Guo, and S.~Luo, ``A high-performance distributed
  file system for large-scale concurrent hd video streams,'' \emph{Concurrency
  and Computation: Practice and Experience}, vol.~27, no.~13, pp. 3510--3522,
  2015.

\bibitem{frazier2018tutorial}
P.~I. Frazier, ``A tutorial on bayesian optimization,'' \emph{arXiv preprint
  arXiv:1807.02811}, 2018.

\bibitem{vikhar2016evolutionary}
P.~A. Vikhar, ``Evolutionary algorithms: A critical review and its future
  prospects,'' in \emph{2016 International conference on global trends in
  signal processing, information computing and communication}.\hskip 1em plus
  0.5em minus 0.4em\relax IEEE, 2016, pp. 261--265.

\bibitem{sutton2018reinforcement}
R.~S. Sutton and A.~G. Barto, \emph{Reinforcement learning: An
  introduction}.\hskip 1em plus 0.5em minus 0.4em\relax MIT press, 2018, ch.~1,
  pp. 1--4.

\end{thebibliography}

\listoftodos
\end{document}